\documentclass[aps,pra,superscriptaddress,twocolumn,amsmath,amsfonts,amssymb,floatfix]{revtex4-1}
\usepackage{enumerate}
\usepackage{mathtools}
\usepackage{amsmath}
\usepackage{graphicx}
\usepackage{epsfig}
\usepackage{sidecap}
\usepackage{hyperref}
\usepackage{color}
\usepackage{subfigure,array}
\usepackage{verbatim}
\usepackage{multirow}

\usepackage{enumerate}
\usepackage{bm}

\begin{document}

\title{Fragmentation of Quantum Fluid in dipolar Bose-Einstein condensate}

\author{Shivam Singh}
\affiliation{Department of Physics, School of Engineering and Applied Sciences, Bennett University, Greater Noida, UP-201310, India}
%\author{Suhail Rashid}
%\affiliation{Department of Physics, IIT Bombay}

\author{Ayan Khan} \thanks{ayan.khan@bennett.edu.in}
\affiliation{Department of Physics, School of Engineering and Applied Sciences, Bennett University, Greater Noida, UP-201310, India}
%\author{Boris A. Malomed}
%\affiliation{Department of Physical Electronics, School of Electrical Engineering, Faculty of Engineering, and Center for Light-Matter Interaction, Tel Aviv University, Ramat Aviv, Tel Aviv P.O. Box 39040, Israel}
\begin{abstract}
	
In this article, we study the dipolar Bosonic quantum fluid. The fluid experiences mean-field, beyond mean-field, and three body interactions. We investigate their competition with dipolar interaction and fragmentation as a result of this competition. We further investigate the elementary excitations and note two distinct dispersion regimes, namely roton-mode and modulational instability. We support our observation by calculating the superfluid fraction and the condensate fraction. 
\end{abstract}

\date{\today}

\maketitle
\section{Introduction}
In the last three decades, we have observed enormous progress in preparation, manipulation, and detection of quantum systems with extreme precision by means of ultra-cold atomic gases, which promises to navigate us toward the new generation of technology, often called `Quantum Technology' \cite{phillips,leggett,stringari1999,stringari2008,bloch,grimm}. 
%By virtue of our knowledge of how to produce pure quantum objects in ultracold atomic systems, we sometimes refer to them as `Quantum Materials'.

%In recent years, the concept of `Quantum Materials' has emerged as a formidable idea that unifies diverse fields of science and engineering and has led to far-reaching applications in atomtronics, quantum simulators, quantum information, and quantum metrology. 
%It is even opening up promising commercial markets in inertial navigation, 
%computation, biomagnetic imaging, mineral exploration, and archaeology \cite{bongs,dowling,boto,snadden,dekker,clauser,meystre,park}. %The western world is already observing a strong commercial 

Of late, some remarkable observations in ultracold gases have ignited several fundamental questions. The observation of liquid-like states, named quantum droplets (QD), in dipolar and binary Bose-Einstein condensates (BEC) is one of the prime examples among contemporary advances in this field \cite{kadau2016observing,cabrera1}. These droplets, first observed in Dysprosium and Erbium gases \cite{kadau2016observing, schmitt2016self}, represent a novel phase stabilization due to the interplay of dipole-dipole interactions and Lee-Huang-Yang (LHY) quantum corrections \cite{wachtler2016quantum, bottcher2019dilute}.
Contrary to the conventional idea, these droplets cannot be associated with the classical concept of liquid, rather they are purely quantum mechanical in nature and manifest quantum fluctuations \cite{kadau2016observing,ferrier2016observation,barbut2}.
%The QDs are the result of two competing interactions, one being the effective mean-field (MF) interaction, and the other being the beyond mean-field (BMF) correction \cite{petrov,sala1,petrov_1d}. The rudimentary theory associated with BMF interaction is based on the Lee-Huang-Yang's (LHY) correction \cite{lee} to the Gross-Pitaevskii equation (GP) \cite{gross,pitae}. 

Further investigations have proven the existence of an exotic phase known as the super-solid phase (SS) in spin-orbit-coupled and dipolar BEC \cite{li, donner, tanzi, bottcher2019transient, chomaz, sohmen1, sohmen2}. The characteristic of this phase is that it combines the properties of solids with those of superfluids (SF). This unique quantum phase of matter possesses lattice ordering, like a solid, and can flow without viscosity! This apparently counterfactual phase featuring antithetical properties of matter can provide us a better understanding of the superfluids and superconductors is precise and quantum many-body dynamics as a whole.

From the perspective of quantum technology, super-solid quantum droplets offer several promising opportunities. Their intrinsic long-range coherence makes them attractive for matter-wave interferometry and quantum sensing. Collective excitation modes and controllable coupling between droplets can be exploited to simulate complex quantum many-body systems, contributing to the development of quantum simulators \cite{barbut2}. Their sensitivity to weak external fields poses the potential for the realization of next-generation precision measurement devices, including gravimeters, inertial sensors, and magnetometers \cite{bongs,park}. The robust coherence of supersolid droplets also makes them a potential platform for studying quantum information transport and coherent state engineering \cite{guo2019low, recati2023supersolidity}.
%This may also lead to a distant implication in realizing the superconducting magnets and sensors \cite{yukalov}.

%Soon after this, experimental groups across Europe were able to observe SS properties in dipolar BEC formed from lanthanide atoms \cite{donner}. In these systems, supersolidity transpires from the competition between two-body scattering and dipolar interaction. The BMF interaction plays the stabilizing role when the other two interactions almost nullify each other \cite{tanzi,bottcher,chomaz}. The experiments were performed of dipolar alkali gases of $^{162}\textrm{Dy}$, $^{164}\textrm{Dy}$ and $^{166}\textrm{Er}$. Over last couple of years the experimental investigation is further extended to understand the emergence and decay of the SS phase at finite temperature has been also been demonstrated experimentally \cite{sohmen1} along with creation of SS phase in two dimensional system \cite{sohmen2}. 

Although early studies treated droplets as coherent single-component systems \cite{baillie2016self, ferrier2016observation}, theoretical and experimental investigations have nonetheless revealed the possibility of fragmentation, which implies spontaneous breaking of coherence into multiple uncorrelated phase substructures. Fragmentation is not merely a spatial separation, but involves a fundamental modification of the quantum state, often associated with the occupation of multiple natural orbitals \cite{cidrim2018vortices}. In dipolar systems, where the anisotropic and long-range nature of interactions introduces a rich landscape of instabilities and pattern formation, the emergence of fragmentation is particularly compelling.

Several mechanisms have been proposed for droplet fragmentation in dipolar BECs, including roton-mode softening, modulational instability, and the role of beyond-mean-field effects in non-equilibrium dynamics \cite{chomaz2018observation, bottcher2019transient}. Recent numerical simulations based on extended Gross–Pitaevskii equations that incorporate the LHY term suggest that fragmentation can arise dynamically during the formation of droplets, especially in the presence of confinement or quench protocols \cite{boudjemaa2023quantum}. However, a comprehensive understanding of the conditions that lead to fragmentation and its experimental signatures remains an open question.

In recent years, we have demonstrated the role of beyond mean-field (BMF) and three-body (3B) interaction in droplet formation for both one-dimensional (1D) and quasi-one-dimensional (Q1D) systems \cite{adusumalli2024quantum, khan2022quantum}. We have also commented on the emergence of a super-solid (SS) like phase due to an additional phenomenological driving force while the mean-field (MF), BMF and 3B interactions keep competing \cite{debnath2022signature}. Furthermore, very recently, we have noted that in a binary BEC the role of 3B and BMF interaction can become more pronounced if we study the Josephson oscillation and self-trapping phases \cite{singh2026josephson}. 
%Using both time-dependent simulations and variational analysis, we identify parameter regimes conducive to fragmentation and quantify the degree of coherence loss. 

We note that in recent years, fragmentation in 1D quantum liquid with LHY correction has already been commented \cite{boudjemaa2023quantum} while the possible existence of SS like phase via the Bogoliubov quasiparticle method in a Q1D system is also reported \cite{turmanov2021oscillations}. Recent experiments also provide us with the knowledge of the superfluid fraction variation with the dipolar interaction strength through the Josephson effect \cite{biagioni2024measurement}.

In this work, we plan to integrate all the earlier efforts in a one comprehensive systematic study for fragmentation of a quantum liquid in dipolar BECs, focusing on the interplay between interaction anisotropy, quantum fluctuations and the three body effect. Specifically, we concentrate on a Q1D system where LHY correction appears in the form of quartic nonlinearity \cite{debnath2021investigation} with dipolar interaction. Additionally, we incorporate the quintic nonlinearity which augurs for 3B interaction \cite{bulgac2002dilute}. We solve the equation of motion for different interaction permutations and carry out a structured study of the dispersion spectrum, superfluid fraction, and condensate fraction as a function of competing interactions. The theoretical structure is described in Sec~\ref{theory}. We present our results in Sec.~\ref{result} and draw our conclusion in Sec.~\ref{conclusion}.

%We expect our results will contribute to a deeper understanding of quantum many-body phenomena in dipolar systems and suggest new pathways for probing fragmented superfluidity and supersolid phase.

\section{Theoretical Model}\label{theory}
A Bose-Einstein condensate with two- and three-body interaction along with quantum fluctuation can be well expressed by the zero temperature mean field and beyond mean field (BMF) theory \cite{turmanov2021oscillations}. The mathematical formulation of inhomogeneous Q1D dipolar Bose-Einstein condensate with non local dipole-dipole interaction can be represented by the following  time dependent Gross-Pitaevskii equation (in natural units) \cite{kumar2015fortran}. 
%\right.\nonumber\\&&\left.	
\begin{widetext}
\begin{eqnarray}\label{Eq. 1D_GP eqn_1}
	i \frac{\partial \phi_{1D} (z, t)}{\partial t}= \left[- \frac{\partial^{2}_{z}}{2} + \frac{\lambda^{2} z^{2}}{2} + \gamma |\phi_{1D}|^{3} + \frac{2 a N_{at} }{d^{2}_{\rho}}|\phi_{1D}|^{2} + G |\phi_{1D}|^{4} + 3 a_{dd} N_{at} \int_{-\infty}^{\infty} V_{dd}^{1D} (|z-z'|) |\phi_{1D}(z', t)|^{2} dz'      \right]   \phi_{1D}(z,t),\nonumber\\
\end{eqnarray}
\end{widetext}

In this equation, we have externally added the beyond-mean-field (BMF) contribution (which appears as a quartic nonlinearity) along with the three-body (3B) contribution introduced via quintic nonlinearity. Here, $\lambda=\omega_z/\bar{\omega}$ with $\omega_z$ being the longitudinal trap frequency and $\bar{\omega}$ is the geometric mean of the transverse and longitudinal trap frequencies and $d_{\rho}$ is the radial harmonic oscillator length scale. The system can be treated as Q1D provided that the transverse confinement is very strong compared to the longitudinal confinement. The $s-$wave scattering length `$a$' is considered as about $113a_B$ or about $6nm$ \cite{kumar2015fortran} and the dipolar scattering length applied here is $a_{dd}=106a_B$. Here, $a_B$ denotes the Bohr radius. Note that both of these interaction strengths are on the same order of magnitude but of opposite polarity. The MF interaction is attractive and the DDI is repulsive. $N_{at}$ describes the number of atoms, $V_{dd}$ takes into account the dipole potential. In addition to these, $\gamma$ and $G$ are noted as the interaction strengths of the quartic and quintic nonlinearities, implying the interaction strength of BMF and 3B interactions, respectively. 

\subsection{Treatment of Dipolar Potential}
To treat the dipolar interaction in Q1D we follow the prescription of Ref.~\cite{kumar2015fortran}. This results in scaling of the dipolar potential such that
\begin{eqnarray}
	&& V_{dd}^{1D}=\frac{2 \pi}{\sqrt{2} d_{\rho}} \left[ \frac{4}{3} \delta(\sqrt{\omega}) \right.\nonumber\\&&\left.+ 2 \sqrt{\omega} - \sqrt{\pi} (1 + 2 \omega)e^{\omega} (1 - erf(\sqrt{\omega}))   \right]
\end{eqnarray}
where $\omega $ = $\left[ Z /(\sqrt[]{2} d_{\rho})   \right]^{2}$, $Z $ = $|z - z'|$. The integral term in Eq.(\ref{Eq. 1D_GP eqn_1}) is solved in momentum space using the convolution identity\cite{muruganandam2009fortran}.

\begin{eqnarray}\label{Eq. Dipolar_Potential}
&&	\int_{-\infty}^{\infty} V_{dd}^{1D} (|z-z'|) |\phi_{1D}(z', t)|^{2} dz'\nonumber\\&& = \frac{4  \pi}{3} \int_{-\infty}^{\infty} \frac{d k_{z}}{2 \pi} e^{-i k_{z} z} \tilde{n}(k_{z}, t)h_{1D} \Big(\frac{k_{z} d_{\rho}}{\sqrt[]{2}} \Big) ,
\end{eqnarray}
where
\begin{eqnarray}\label{density_mom_space}
	\tilde{n}(k_{z}, t)&=&\int_{\infty}^{\infty} e^{i k_{z} z} |\phi_{1D} (z, t)|^{2} dz, \nonumber\\
    	\tilde{n}(\boldsymbol{k}_{\rho} )&=&\int e^{i \boldsymbol{k}_{\rho}. \rho} |\phi_{2D} (\rho)|^{2} d \rho  = e^{- k_{\rho}^{2} d_{\rho}^{2}/4}, k_{\rho} = \sqrt[]{k_{x}^{2} + k_{y}^{2}},\nonumber\\
\end{eqnarray} 
and,
\begin{eqnarray}\label{h1d}
	&& h_{1D} (\zeta) = \frac{1}{(2 \pi)^{2}} \int d \boldsymbol{k}_{\rho} \Big[ \frac{3 k_{z}^{2}}{\boldsymbol{k^{2}}} - 1 \Big] |\tilde{n}(\boldsymbol{k}_{\rho}^{2})|^{2} \nonumber\\&&= \frac{1}{2 \pi d_{\rho}^{2}} \int_{- \infty}^{\infty} du \Bigg[ \frac{3 \zeta^{2}}{u + \zeta^{2}} - 1 \Bigg] e^{-u}, \zeta = \frac{k_{z} d_{\rho}}{\sqrt[]{2}}
\end{eqnarray}

One can now write down Eq.(\ref{Eq. 1D_GP eqn_1}) in a simpler form as
\begin{eqnarray}\label{Eq. 1D_Gp_eqn_2}
&&	i \frac{\partial \phi_{1D} (z, t)}{\partial t}= \left[- \frac{\partial^{2}_{z}}{2} + \frac{\lambda^{2} z^{2}}{2} + g_{0} |\phi_{1D}|^{2} + g' |\phi_{1D}|^{3} \right.\nonumber\\&&\left.+ g_{dd} \phi_{dd} + g_{30} |\phi_{1D}|^{4}   \right] \phi_{1D}(z,t)
\end{eqnarray}
where, $\phi_{1D}(z, t)$ represents the mean field wave function of the condensate, $g_{0}$, $g_{30}$ are the coefficients of two-body mean-field (MF) and three-body (3B) interaction strengths, respectively. The parameter $g'$ denotes the contribution from quantum fluctuation or the BMF contribution. $g_{0}$ and $g_{dd}$ are contact and dipolar interaction strengths where $g_0=4\pi a $ and $g_{dd}=3 a_{dd}$ in units of Bohr-radius, respectively \cite{kumar2015fortran}. Furthermore, the long-range dipole-dipole interaction (DDI) is represented by the potential $\phi_{dd}(z, t) = \int dz' V_{dd} (z - z') |\phi_{1D}(z', t)|^{2}$. 

The contribution of quantum fluctuation is characterized by \cite{baillie2016self} $$g' = \frac{32  g_{0} \sqrt{a^{3}}}{3 \sqrt{\pi}} \Bigg[ 1 +\frac{3}{2} \epsilon_{dd}^{2}  \Bigg]$$.
Here, $\epsilon_{dd}$ is the ratio between the dipolar and s-wave scattering length as $\epsilon_{dd} = a_{dd}/a$. 3B interaction strength is illustrated as $g_{30} = 3g_0a^{3} [d_{1} + d_{2} \tan(s_{0} \ln(|a|) + \pi /2)]$\cite{efimov1970energy, bulgac2002dilute}. Here $s_{0}$ is the dimensionless constant where $s_0=1.0064$. The universal constants $d_{1}$ and $d_{2}$ is defined as $1.23$, $- 3.16 $, respectively \cite{braaten2002dilute}. %Applying these parameter values, the magnitude of the interaction strengths works out as, 
%$g_{0} = 75.1431$, $g' = 161.7409$, $g_{30} = -1.57577 \times 10^{-5}$, $g_{dd} = 16.827835 $ respectively. All the numerical values are in the units of Bohr-radius scaled by harmonic oscillator length scale ($d_{\rho}$). %All the needed calculation regarding the DDI has been done and very well described in Ref.~\cite{kumar2015fortran}.  
\subsection{Elementary Excitations}  
The spatial period of the emerging density modulations in a dipolar BEC can be directly theorized from the dispersion relation of its elementary excitations. In dipolar Bose–Einstein condensates (DBECs), the anisotropic and long-range nature of the DDI necessitates a modification in the Bogoliubov spectrum in comparison to the contact-interaction picture. However, under suitable interaction strengths and trapping conditions, the excitation spectrum may develop a roton-like minimum at a finite momentum. Therefore, the dispersion curve of elementary excitations provides a rudimentary way to investigate the spatial period of density modulations.

Usually in a uniform superfluid, excitations at low momentum are sound waves (phonons). However, at higher, finite momenta, the energy required to excite the fluid dips, creating a local minimum in the energy-momentum dispersion relation. This localized excitation is called a roton.
In a DBEC, the long-range and anisotropic dipole-dipole interactions can cause this roton minimum to drop closer and closer to zero energy. When the energy of the roton mode completely softens or approaches zero, the system undergoes a roton instability.
This instability triggers the phase transition from a uniform superfluid to a super-solid, causing the quantum gas to spontaneously break the translational symmetry and freeze into a periodic crystalline array of droplets while remaining perfectly phase-coherent.

 The dispersion relation, in terms of constant background density ($n_{0}$) and  uniform phase ($\theta$), can be obtained by initializing the form of the ground state wavefunction as \cite{turmanov2021oscillations}: 
\begin{equation}
	\phi_{0} = \sqrt{n_{0} } e^{i \theta t},
\end{equation}
where 
\begin{eqnarray}
	\theta &=& n_{0} \Big(  g_{0}   + g_{30}  n_{0} + g' n_{0}^{1/2}  + g_{dd} \int_{- \infty}^{\infty} V_{dd} (Z) dz \Big)
\end{eqnarray}

The Fourier transform of the DDI potential ($V_{dd}(Z)$) is calculated numerically by using Gauss–Legendre quadrature in momentum space, resulting the quasi one-dimensional dipole kernel ($ \tilde{V}_{dd}(k)$) as noted here \cite{kumar2015fortran},
\begin{eqnarray}
	 \tilde{V}_{dd}(k) &=& \frac{2}{3 d_{\rho^{2}}} \int_{0}^{\infty} \Big(  \frac{3 k_{z}^{2}}{k_{\rho}^{2} + k_{z}^{2}} - 1  \Big) k_{\rho} e^{-k_{\rho}^{2} d_{\rho}^{2}/2} dk_{\rho} 
\end{eqnarray}

One can now obtain the elementary excitation spectrum by executing a Bogoliubov linear stability evaluation around the uniform stationary solution. In the context of DBECs,we have performed the Bogoulibov analysis by taking into account the different interaction competitions which includes the MF, 3B and BMF interactions. The resulting dispersion relation can now be written as\cite{turmanov2021oscillations},
\begin{eqnarray}
	\Omega(k)&=&\Bigg[k^{2} \Bigg(   \frac{k^2}{4} - n_{0}  \Big(   g_{0} + 2 g_{30} n_{0} + \frac{3}{2} g'  n_{0}^{1/2} + g_{dd} \tilde{V}_{dd}(k)     \Big)   \Bigg) \Bigg]
	\nonumber\\\label{dispersion_relation}
\end{eqnarray}

\subsection{Superfluid \& Condensate Fraction}
In a dipolar Bose-Einstein Condensate (BEC), anisotropic, long-range dipole-dipole interactions significantly modify both the superfluid fraction and the condensate fraction. These properties control the system's ability to transition into exotic phases like super-solids and quantum droplet arrays. In this subsection, we summarize the methodology to calculate the superfluid fraction ($f_s$) and condensate fraction ($f_c$).
\subsubsection*{Superfluid Fraction ($f_{s}$)}
To calculate $f_s$ we borrowed the theory of Leggett's upper bound, where it is suggested that for a given direction, the superfluid fraction cannot exceed the inverse spatial average of the inverse local density. Mathematically, it is defined as \cite{pitaevskii2016bose}:
\begin{equation}
	f_s =  \frac{(2 L)^2}{\int_{-L}^{L}dx \text{ } \tilde{n}(x) \int_{-L}^{L} \frac{dx}{\tilde{n}(x)} }\label{SF_eqn}
\end{equation} 
Where , $x$ is the coordinate along which the droplet array is formed, $\tilde{n}(x)$ column density, and $2 L$ is the length that surrounds the central region. In general, for uniform superfluid at zero temperature $f_s=1$, however, in a dipolar BEC, when the density increases enough to form super-solids or droplet arrays, translational symmetry is broken suppressing the zero temperature superfluid fraction i.e., $f_s<1$. 

\subsubsection*{Condensate Fraction($f_{c}$)}
The condensate fraction ($f_c$) describes the proportion of particles occupying the exact same macroscopic quantum ground state. In dipolar BEC the dipole interaction increases quantum fluctuations resulting depleted fraction even at zero temperature.

We calculate the condensate fraction from the 
one-body density matrix (OBDM), introduced by Penrose and Onsager \cite{penrose1956bose}.
For a Bose field operator $\hat{\Psi}(x)$, the OBDM defined as,
\begin{equation}
	\rho_1(x,x') =
	\langle \hat{\Psi}^\dagger(x)\,\hat{\Psi}(x') \rangle .
	\label{eq:obdm_def}
\end{equation}

The OBDM measured the quantum coherence between two spatial points $x$
and $x'$. Its spectral decomposition is obtained by the eigenvalue problem yielding, 
\begin{equation}
	\int dx' \, \rho_1(x,x') \, \phi_j(x') = \lambda_j \, \phi_j(x).
	\label{eq:obdm_eigen}
\end{equation}
Here, $\{\phi_j(x)\}$ are the natural orbitals and
$\{\lambda_j\}$ are their occupation numbers.

The condensate fraction is then defined as \cite{penrose1956bose},
\begin{equation}
		f_c = \frac{\lambda_0}{N}\label{CF_eqn}
\end{equation}
where $\lambda_0$ is the largest eigenvalue of the OBDM and $N = \int dx \, n(x)$ is the total number of particles.

%This criterion is known as the \emph{Penrose--Onsager criterion} for
%Bose-Einstein condensation.

\section{Results}\label{result}
%\subsection{Discussion}
Here we present our results which contain the calculation of the ground state wave-function for dipolar BEC by numerically solving Eq.(\ref{Eq. 1D_Gp_eqn_2}). The numerical solution also takes into account the effect of BMF and 3B interaction. 
As a trial wave function we have used the recently reported analytical solution for droplet state in binary BEC \cite{debnath2021investigation}.
%\begin{eqnarray}\label{eq: inialize wavefunction}
%	\psi(z)&=&\frac{1 + 12 \mu}{1+\sqrt{12 \mu} \hspace{0.5mm} cosh(z)}
%\end{eqnarray}
To solve Eq.(\ref{Eq. 1D_Gp_eqn_2}), we use imaginary time program method. In the numerical calculation we have considered the space mesh point $N = 2048$, space step, $dz = 0.05$, time step, $dt = 0.005$ and number of atoms, $N_{at} = 1000$. 

After through analysis of each of the Hamiltonian parameters, we choose to present the following variation: (i) Variation of dipolar interaction with fixed MF, BMF and 3B interaction at fixed particle number; (ii) Variation of MF interaction with DDI, BMF and 3B interaction at fixed particle number; (iii) Varying particle number with fixed MF, DDI, BMF and 3B interaction; (iv) Varying particle number in absence of dipolar interaction with fixed MF, BMF and 3B interaction. For each case we solve Eq.(\ref{Eq. 1D_Gp_eqn_2}) numerically and then we study the dispersion mechanism, condensate fraction, and superfluid fraction to comment on the nature of the quantum fluid. Although several other combinations of interaction strengths have also been explored in our study, the above cases serve as the primary focus of our current discussion. This is because they clearly illustrate the dominant physical mechanisms arising from the competition between different interaction terms and show the fragmentation characteristics.

%\subsection{Results}
%In preceding sections we have discussed theoretical model of dipolar Gross-Pitaevskii equation (DGPE) in which provides the details of non linear terms and corresponding interaction strengths. Here we plan to examine our results for all competing strength. In particular, we have taken scattering length $a = 113$$a_{b}$ and dipolar length $a_{dd} = 106$$a_{b}$ for trap parameter, $\lambda = 1$ radial Gaussian width, $d_{\rho} = 1$   and for the fixed number of atoms, $N_{at}$ and fixed space ($dz$) and time steps ($dt$). We have seen for different interaction strength considerable remarks can be established. 

\begin{figure*}
	\centering
	\includegraphics[width = 0.90\linewidth]{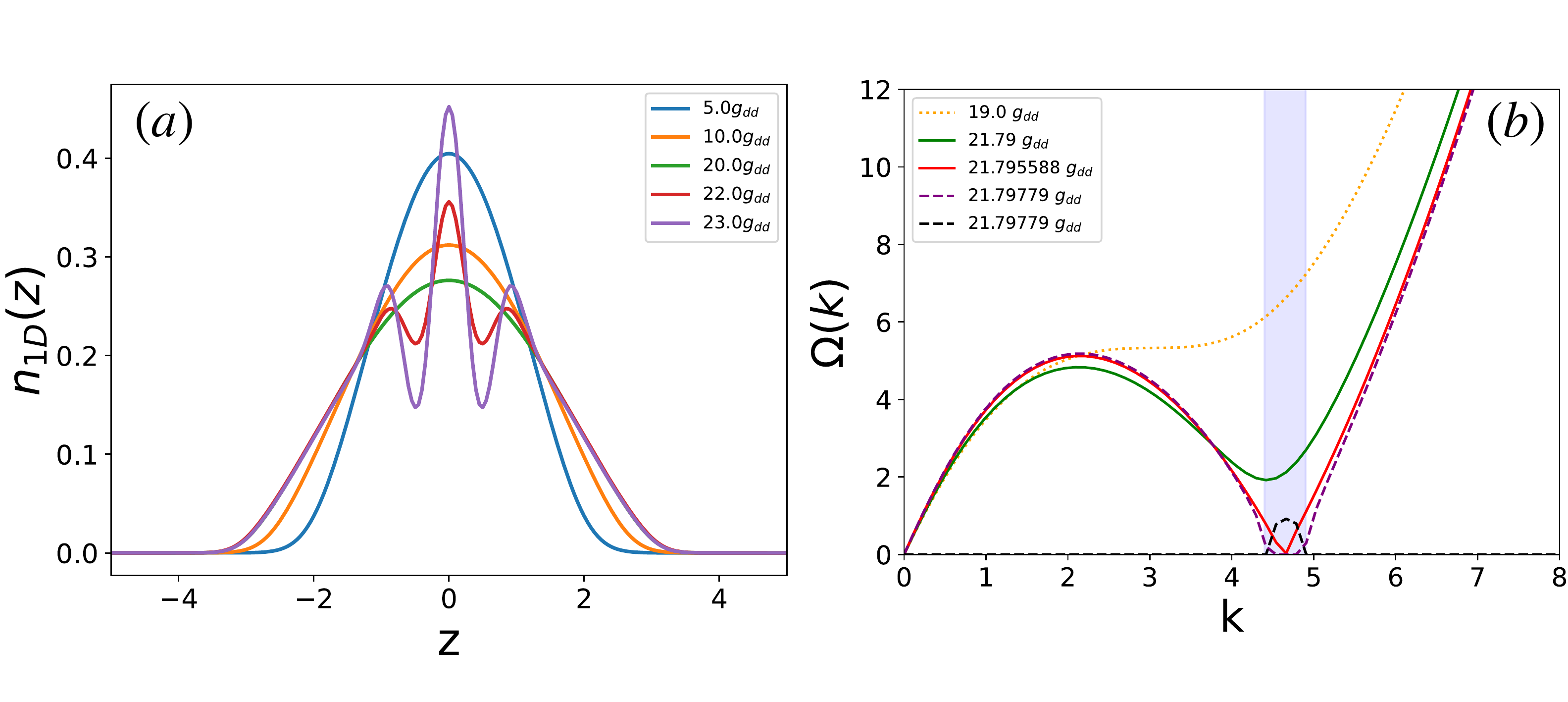}  %Case-1
	\caption{ (color online) ($a$) The ground state density profile $n_{1D}(z)$ with spatial coordinate $z$ is reported here. In this case other interactions except DDI is considered to be fixed. The magnitude of MF interaction strength is set at $g_0=4\pi a$ and it is being attractive in nature, the BMF strength is $g'/g_0=60\sqrt{a^3}$ and $g_{30}/g_0=3a^3[1.23-3.16\tan(\ln(|a|))+\pi/2]$ (assuming $s_0\sim1$). Both the BMF and 3B interactions are repulsive. The $g_{dd}$ is varied by increasing its strength from $5.0g_{dd}$ to $23.0g_{dd}$. The particle number $N_{at}$ is set at $1000$. ($b$) Shows the dispersion relation by accounting the Eq.\ref{dispersion_relation} along with background density ($n_{0}$) values and the shaded blue region illustrates the region of instability.}\label{case1_fig}
\end{figure*}

\subsection*{Variation of Dipolar Interaction}
In this section, we solve Eq.(\ref{Eq. 1D_Gp_eqn_2}) for different dipolar strength. Typically we use attractive mean field interaction, repulsive BMF and 3B interaction as well as repulsive dipolar interaction ($g_{0} < 0$, $g' > 0$, $g_{dd} > 0$, $g_{30} > 0$). We vary the dipolar interaction from $5.0$ to $20.0$ in units of $g_{dd}$.

The numerical solution provides us with a localized solution that gets gradually fragmented as we increase the dipolar strength. We have described this feature in the left panel of Fig.~\ref{case1_fig}. The blue, orange and green solid line describes the localized solution for a fixed number of particles with dipolar strength gradually increased from $5g_{dd}$ to $20g_{dd}$. However, a further increase in dipolar interaction shows fragmentation of the quantum fluid (as described by the red and purple solid lines). Even though we report the fragmentation phenomena as a function of dipolar interaction, but we must also note from our investigation that the non-zero BMF interaction is essential to yield fragmentation. Hence, we can safely conclude that the fragmentation is supported due to the presence of BMF interaction. 
 
To investigate the fragmentation mechanism more closely, we study the dispersion curve of the system which is reported in the right panel of Fig.~\ref{case1_fig}. Our density plot suggested a possible fragmentation between $20g_{dd}$ and $22g_{dd}$. Specifically, it can be seen that the localized nature of the density profile begins to disappear around $\sim 21.0g_{dd}$. At this stage, the previously localized density profile starts to get fragmented with multiple peaks appearing in the spatial density distribution. This structural change in the density profile is fully supported by the nature of excitation spectrum shown in the right panel. At interactions in the vicinity of fragmentation we observe that the dispersion curve starts bending (please see yellow dotted line and green solid line). As the density dispersion starts manifesting fragmentation, the dispersion curve also shows characteristics of roton mode. 
In particular, the roton minimum becomes significantly pronounced as the dipolar interaction strength increases from $21.0g_{dd}$ to approximately $21.79g_{dd}$. Further increase leads to the region of instability where a section of dispersion frequency becomes complex. We observe a region of instability for $21.79779g_{dd}$ where some part of $\Omega(k)$ is complex. The purple dashed line describes the real part of $\Omega(k)$ while the black dashed curve denotes the imaginary part of $\Omega(k)$. When DDI is $21.795588 g_{dd}$ we can observe the roton minima (red solid line) where the elementary excitation frequency is zero at finite momentum.  
\begin{comment}
\begin{figure}
	\centering
	\includegraphics[width = 0.95\linewidth]{energy_rms.pdf} %G0_-ve_GDD0_G30_LHY_+ve_ED_var_GDD0.pdf
	\caption{ Variation of the energy per particle and the r.m.s. size as a function of the dipolar interaction strength. The left panel shows the behavior of the energy per particle, while the right panel represents the corresponding variation of the r.m.s. size with increasing dipolar strength. All the remaining parameters and constants are kept identical to those used in Figure~\ref{case1_fig}. }
	 \label{fig:CaseI_-g0_-ve_g'_g30_+ve_Energy_RMS}
\end{figure}
\end{comment}

The softening of the roton mode indicates that the system is approaching a regime where density modulations become energetically favorable. After further increasing the dipolar interaction strength, the roton minimum continues to go deeper and eventually reaches zero energy, signifying the critical momentum beyond which roton instability emerges. At this point, the condensate becomes unstable against the formation of periodic density modulations, which refers to a state that simultaneously exhibits superfluid coherence and spatial density ordering which defining characteristics of a super-solid phase.
\begin{comment}
In Fig.\ref{fig:CaseI_-g0_-ve_g'_g30_+ve_Energy_RMS}, we have mentioned  energy per particle ($E / N $) and the root-mean-square ($\sqrt{\langle z^{2} \rangle}$) size of the corresponding density solution. It shows that while the energy per particle is linearly decreasing with increasing dipolar strength, the root mean square size of the clustered atoms are increasing pretty rapidly in the initial period while stabilizing or saturating near $20g_{dd}$.
\end{comment}

\begin{figure}
	\centering
	\includegraphics[width = 0.90\linewidth]{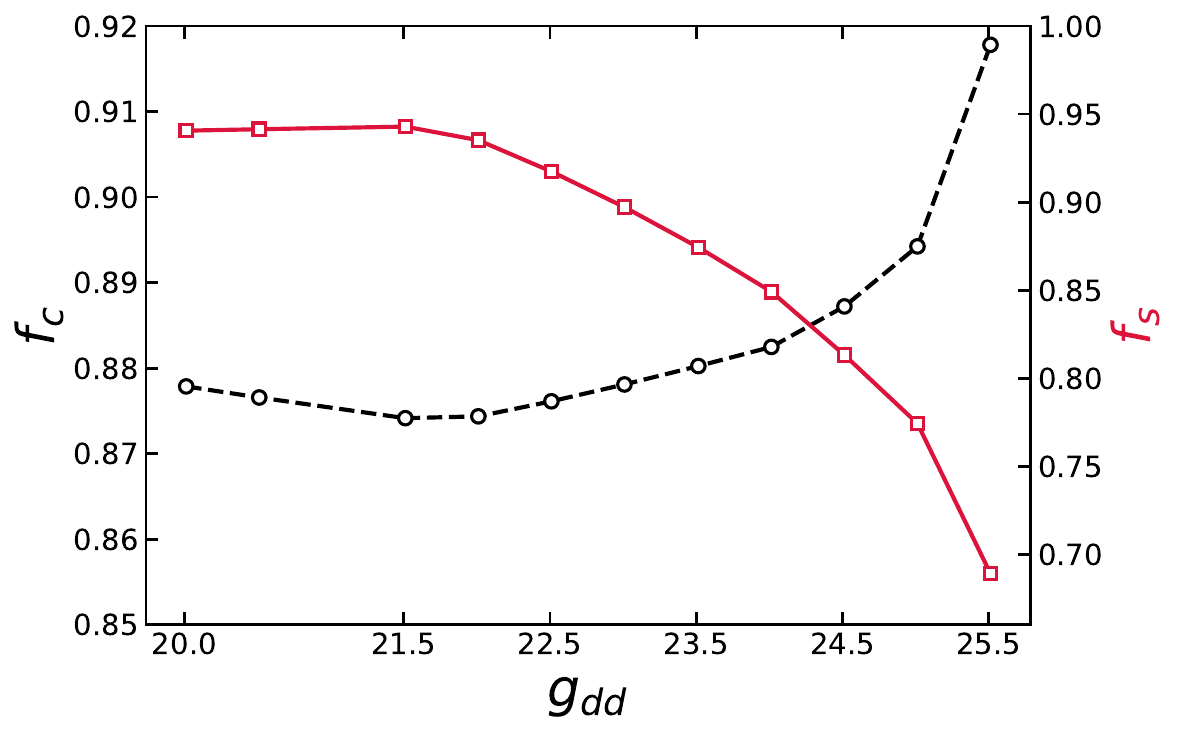}
	\caption{(Color online) Superfluid fraction (red dashed curve) and condensate fraction (black dashed curve) as functions of the dipolar interaction strength $g_{dd}$ is reported here. The results are obtained from the solution presented in Fig.~\ref{case1_fig}. The figure clearly illustrates the distinct dependence of the superfluid and condensate fractions on the dipolar interaction.}\label{Sf_CF_case1}
\end{figure}

Next we calculate the superfluid fraction ($f_s$) and condensate fraction ($f_c$) from our obtained solution and report it in Fig.~\ref{Sf_CF_case1}. The $x$-axis denotes the variation of dipolar strength, while the left $y$-axis describes $f_c$ and right $y$-axis provides the scale for $f_s$. The black dotted line corresponds to $f_c$ while the red solid line corresponds to $f_s$. We observe a roughly $28\%$ decrease in the superfluid fraction while $\sim5\%$ growth in the condensate fraction as we increase the dipolar interaction strength.

%In dipolar quantum gases, increasing the strength of the dipole-dipole interaction (DDI) leads to the simultaneous decrease of the superfluid fraction and increase of the condensate fraction as observed here. 
This counterintuitive behavior occurs because strong DDI drives the system toward crystallization, producing a roton instability that breaks translational symmetry while leaving particles locally condensed \cite{he2025dipolar}.
As the dipolar interaction strength increases, it causes a prominent ``roton minimum'' in the excitation spectrum of the quantum gas (as noted in Fig.~\ref{case1_fig}). As a result, it becomes energetically favorable for the system to develop a periodic density modulation. The quantum fluctuations act to stabilize these structured states. Within these self-bound or structured phases (like quantum droplets or super-solids), atoms become highly localized, which drives a significant reduction in quantum depletion, thereby boosting the actual fraction of atoms in the zero-momentum condensate state. In contrast, superfluid fraction is unity in an ideal superfluid while zero for standard crystals. The suggestive spontaneous crystallization of the quantum fluid, as noted through our density distribution and dispersion analysis obtains further support through the observation of supression of the superfluid fraction. Here the system's mass is pinned to the underlying lattice due to the possible emergence of SS phase \cite{bombin2017dipolar, dong2026characterizing}. The suppression of superfluid fraction is already observed in dipolar condensate and our result qualitatively agrees to the experimental observation \cite{biagioni2024measurement}.

\begin{comment}
\begin{figure}[H]
	\centering
	\includegraphics[width = 0.83\linewidth]{G0_-ve_GDD0_G30_LHY_+ve_ED_var_G0.pdf}
\end{figure}
\end{comment}

\begin{figure*}
	\centering
	\includegraphics[width = 0.90\linewidth]{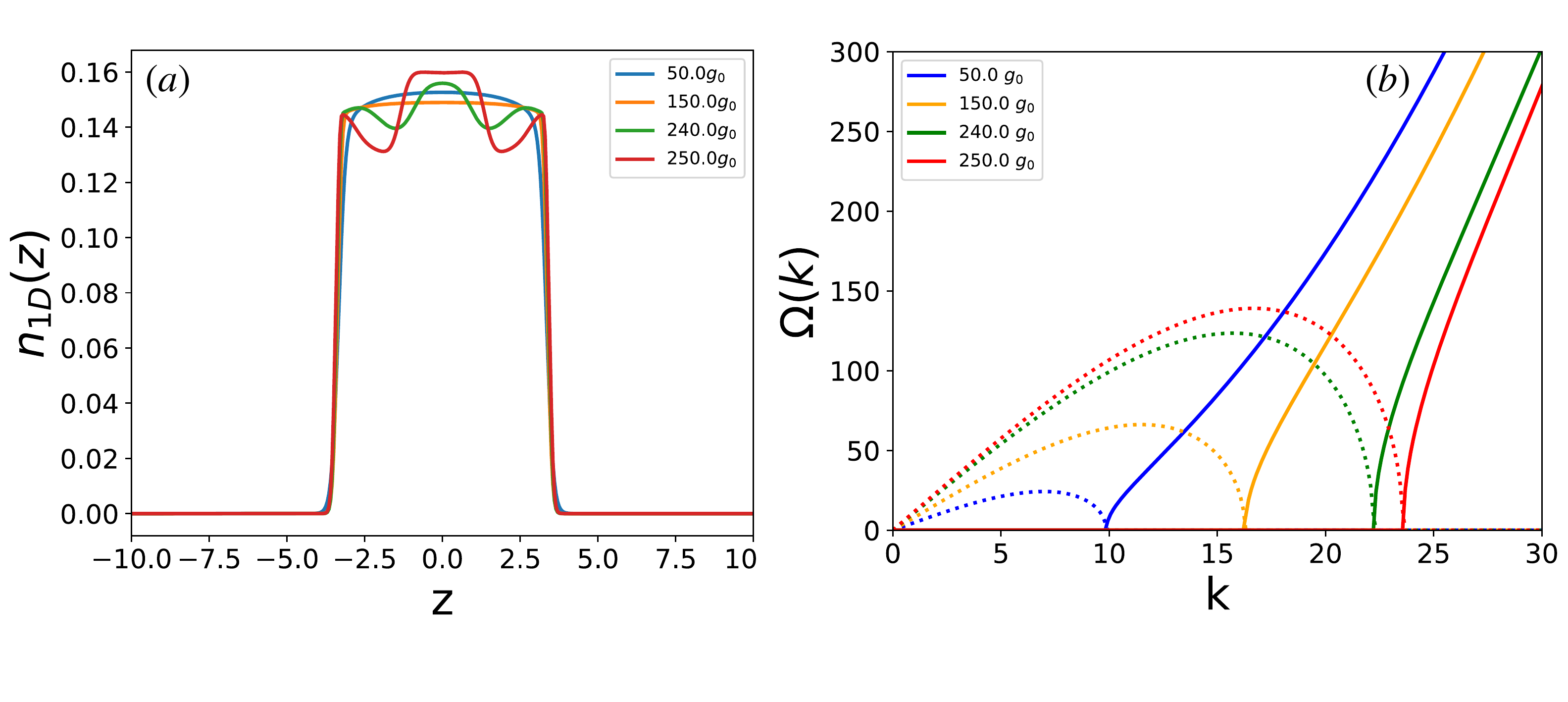}  %Case-1
	\caption{(color online) ($a$) The ground state density profile $n_{1D}(z)$ with spatial coordinate $z$ is depicted here. ($b$) The corresponding dispersion relation is exhibited here. Here, MF interaction is varied from $-50g_0$ to $-250g_0$. BMF strength used as $g'/g_0=60\sqrt{a^3}$ and $g_{30}/g_0=3a^3[1.23-3.16\tan(\ln(|a|))+\pi/2]$ (assuming $s_0\sim1$). Both the BMF and 3B interactions are attractive in nature. The dipolar interaction is fixed at $g_{dd}=3a_{dd}$ and repulsive in nature. The particle number is set at $1000$.} \label{Case-2_den-desp}
\end{figure*}

\subsection*{Variation of Mean-field Interaction}
In the preceding section, we have studied the fragmentation for localized profiles. Here, we summarize our observations for flat-top solutions or quantum droplets. We observe the fragmentation phenomena of the droplets by changing the attractive mean-field interaction while keeping the BMF, dipolar and 3B interactions fixed. Interestingly, we observe the fragmentation when the BMF interaction is negative (in contrast to the common description in Q1D) and dipolar interaction as repulsive. The 3B interaction is also taken as attractive. Here, it must be noted that in 1D system, we have observed that the both mean-field and BMF interactions are treated as attractive while studying the Josephson dynamics \cite{wysocki2024josephson} contrary to the usual description of repulsive mean-field and attractive BMF interaction \cite{petrov_1d}. This liberty is possible in ultracold systems due to the unprecedented control over interaction through external magnetic fields. Here, the presence of the repulsive dipolar interaction counter balances the other attractive interaction strengths and offers flat-top solution. We observe stable droplet solution and gradual fragmentation by tuning the attractive mean-filed interaction. We observe that non-zero contribution of BMF interaction is essential for this droplets to emerge.  

%Now, we are interested in to take account of 2B, 3B and BMF interaction negative and DDI positive ($g_{0} < 0$, $g' < 0$, $g_{dd} > 0$, $g_{30} < 0$). Accordingly,  significant amount of changes in density solution can be seen after the varying different interaction strengths. Particularly, all the interactions plays a decent role in the droplet formation.  
It is well understood from the Fig.~\ref{case1_fig}(left) that the nature of density solution will be localized for relatively low attractive interaction strength (like, $-g_{0}$ to $-10 g_{0}$), however, if the mean-field attractive interaction get stronger, we start observing the droplet formation (till about  $-230 g_{0}$) as described in Fig.~\ref{Case-2_den-desp} and further we observe the fragmentation (say at $-240 g_{0}$). In the Fig.~\ref{Case-2_den-desp} the blue and orange solid line correspond to $-50g_0$ and $-150g_0$ while the green and red solid line depicts $-240g_0$ and $-250g_0$ respectively.

\begin{comment}
\begin{figure}
	\centering
	\includegraphics[width = 0.83\linewidth]{G0_G30_LHY_-ve_GDD0_+ve_ED_var_LHY.pdf}
\end{figure}

\begin{figure}[h]
	\centering
	\includegraphics[width = 0.95\linewidth]{G0_G30_LHY_-ve_GDD0_+ve_ED_var_G0.pdf}
	\caption{Variation of the energy per particle and the r.m.s. size as a function of the two-body (2B) interaction strength. All other interaction parameters and fixed values are taken to be the same as those used in Figure~\ref{Case-2_den-desp}. Left plot showing the enegy per particle with 2B coupling strength. While right figure is the variation of r.m.s with 2B interaction strength. }
	\label{fig:CaseII_-g0_g'_g30_-ve_gdd_+ve_Energy_RMS}
\end{figure}
\end{comment}
To probe the flat-top and fragmented states more closely, we calculate the dispersion relation from Eq.(\ref{dispersion_relation}), which is illustrated in the right panel of Fig.~\ref{Case-2_den-desp}. Structurally, Eq.~\ref{dispersion_relation} is analogous to the Andreev model \cite{andreev2013self}, where the excitation spectrum exhibits competition among quantum pressure, short-range interactions, and polarization effects.
The dispersion relation clearly exhibits the modulational instability (MI) where we can see the side lobe generation from the complex $\Omega(k)$. Here, we note that modulational instability is already been observed in the Q1D optical lattices \cite{otajonov2025modulational}. The imaginary $\Omega(k)$ is commonly known as gain parameter in MI terminology. The area of the side-lobes increases as we move from flat-top droplets to fragmented droplets.

In the context of nonlinear physics, the area under the MI side lobe spectrum is directly proportional to the instability growth rate and the depth of the density modulation. As the droplets fragment further, the modulation depth increases and the gain spectrum broadens. Consequently, the total spectral area contained in the side lobes increases as energy redistributes into the newly formed, fragmented droplet states. We can also note from the figure that the critical wavevector ($k_c$) beyond which the $\Omega(k)>0$ is progressively increasing as we increase the mean-field interaction. 
\begin{figure}
	\centering
	\includegraphics[width = 0.90\linewidth]{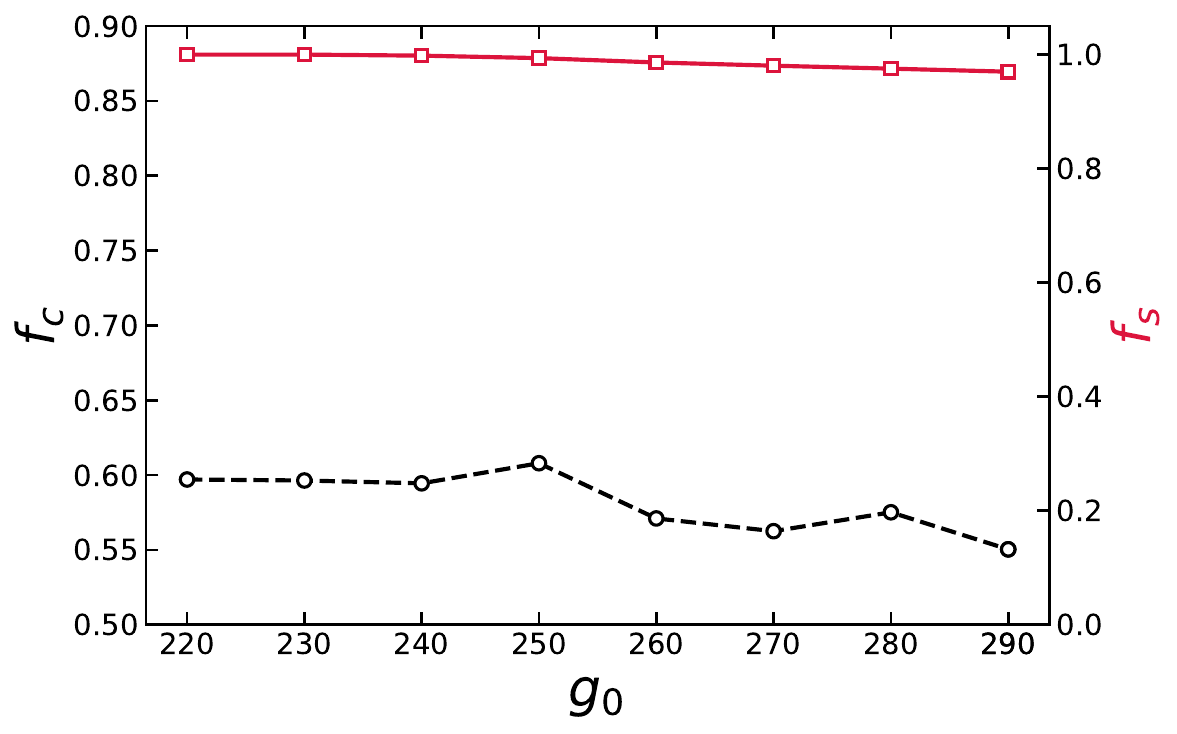}
	\caption{(color online) The superfluid fraction and condensate fraction are depicted here. The red solid and black dashed lines describes $f_s$ and $f_c$ respectively for different values of 2B interaction strength ($g_{0}$).}\label{Sf_CF_case-II}
\end{figure}

%sThe energy per particle and the root-mean-square size ($\sqrt{\langle z^{2} \rangle}$) of the droplets show a behavior similar to that of the previous. However, careful observation reveals that the $\sqrt{\langle z^{2} \rangle}$ is larger (see Fig.~\ref{fig:CaseII_-g0_g'_g30_-ve_gdd_+ve_Energy_RMS}) compared to the reported size in Fig.~\ref{fig:CaseI_-g0_-ve_g'_g30_+ve_Energy_RMS}, which is understandable as we were looking at the localized structure in the previous case while we are now working with the quantum droplets. The energy per particle data reveal the stronger binding energy of the particles compared to the localized ones. 

%The analysis also explicate that as the 2B interaction strength is increased, the energy per particle decreases monotonically, indicating a tendency of the system toward a more energetically favorable configuration. At the same time, the corresponding r.m.s. size shows an increasing behavior, implying that the droplet becomes more spatially extended. This suggests that the enhanced repulsive contribution effectively balances the attractive components, leading to a successively extension of the condensate. 

In Fig.~\ref{Sf_CF_case-II} the superfluid and condensate fraction variation is depicted for variation of MF interaction. All other parameters remain fixed like other plots in this section. The solid red line describes the superfluid fraction while the black dotted line depicts the condensate fraction. We observe superfluid fraction remains nearly unity while condensate fraction is relatively low and slightly reduces ($\sim 8\%$) with increase in the mean-field interaction. This discrepancy arises because quantum fluctuations deplete the zero-momentum state, while superfluidity is primarily dictated by the system's overall phase coherence and density correlations. So one can safely conclude that the quantum fluid remains superfluid even after fragmentation while condensate fraction drops marginally.

\subsection*{Role of number of particle ($N_{at}$)}
In our previous study we have seen that number of atoms ($N_{at}$) also plays significant role for stable droplet formation \cite{debnath2022dropleton}. Therefore, we extend our investigation towards the role of particle number as well. In precise, we study the role of particle number in presence and absence of dipolar interaction. 

\begin{figure*}
	\centering
	\includegraphics[width = 0.95\linewidth]{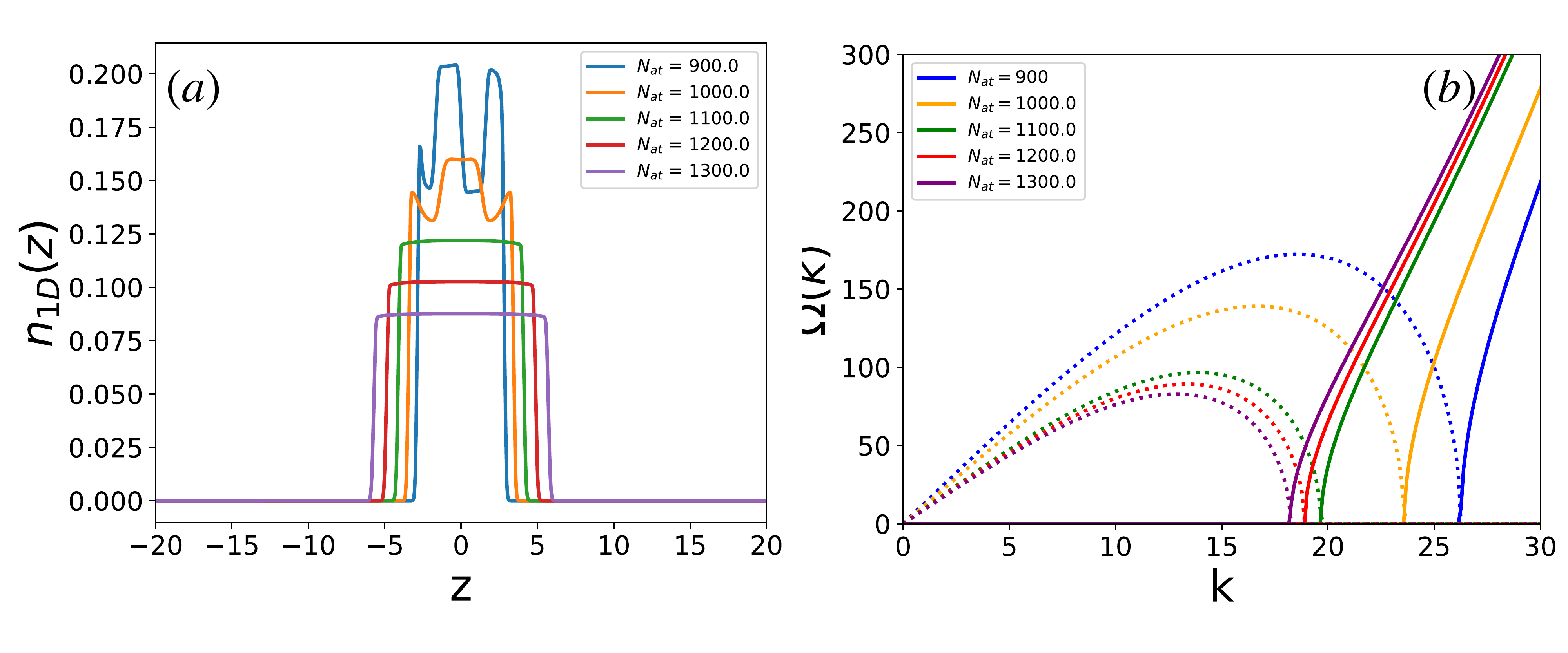}
	\caption{(color online) ($a$) Density profiles of the dipolar Bose gas with varying atom number is presented here. The atom number varies from $900$ to $1300$. In this case, the two-body (2B) interaction strength is fixed at $-250g_{0}$. The contributions from the 2B, 3B, and BMF correction are considered to be attractive (BMF strength used as $g'/g_0=60\sqrt{a^3}$ and $g_{30}/g_0=3a^3[1.23-3.16\tan(\ln(|a|))+\pi/2]$ (assuming $s_0\sim1$).), whereas the DDI is taken to be repulsive. ($b$) The corresponding elementary excitation curve is depicted.}\label{G0_250_var_NATOMS}
\end{figure*}

\begin{figure*}
 	\centering
 	\includegraphics[width = 0.90\linewidth]{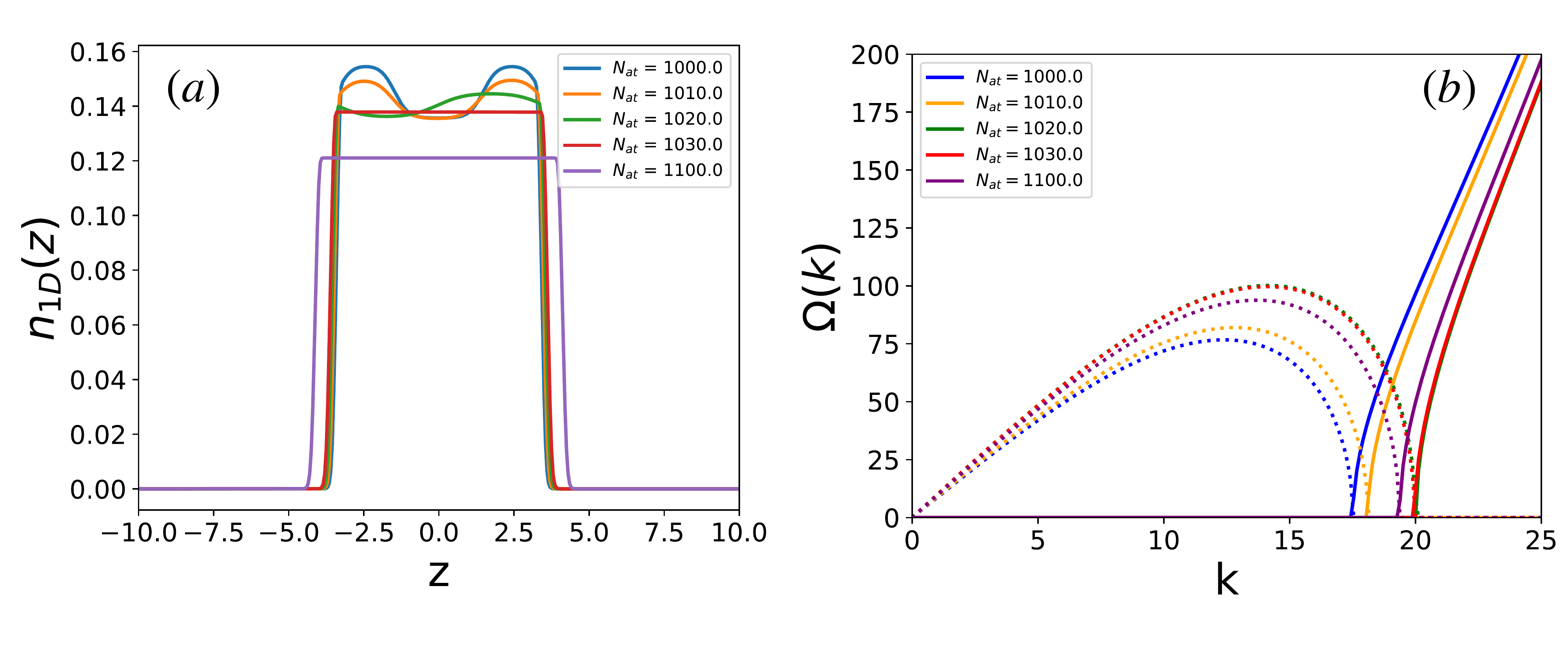}
 	\caption{(color online) ($a$) Showing the density profile, $n_{1D}(z)$, as a function of the spatial coordinate $z$. 2B, 3B, and BMF interaction strengths are taken to be attractive and their numerical values are same as the values used in Fig.~\ref{G0_250_var_NATOMS}. The only difference here is that, we have switched off DDI i.e., $g_{dd} = 0.0$. The number of atoms, $N_{at}$, varies from $1000$ to $1100$. ($b$) Description of the modulational instability arising from the dipolar Bose system. \label{G0_250_DDI_0.0_var_NATOMS}}
 \end{figure*}

For this purpose, we consider the situation where DDI is repulsive, and rest of the interactions are attractive in nature. Applying original interaction strength of 2B, 3B, BMF and DDI and see how number of atoms effects the density solution. Impetus of density solution for different number of atoms is depicted in Fig.\ref{G0_250_var_NATOMS}. We observe that the density evolves from localized to flat-top to fragmented droplets as a result of an increase in the particle number. Therefore, by fixing the 2B interaction strength at $-50g_{0}$ and gradually increasing the number of particles ($N_{at}$), the condensate undergoes a significant change in its density distribution. Specifically, our system evolves from localized density profile to flat-top density profile. This indicates that the increase in particle number enhance the effective nonlinear interactions, which promotes the formation of a self bound droplet with nearly uniform central density. Furthermore, as the 2B interaction strength is increased further, we observe the emergence of a fragmented density nature at ($-250g_{0}$). This fascinating behavior motivated us to investigate whether the fragmentation persists with an increase in the number of particles. To address this question, we considered two different scenarios: one in the presence of DDI (see Fig.~\ref{G0_250_var_NATOMS}) and the other in the absence of DDI (see Fig.~\ref{G0_250_DDI_0.0_var_NATOMS}). Our findings suggests that, irrespective of the presence of DDI, increasing $N_{at}$ gradually suppresses the fragmented density profile. Eventually, the fragmented solution disappears completely in both cases. In our analysis we observe that for $N_{at}\sim1100$ the quantum fluid starts showing signature of droplet formation. The same signature we observe for both the density distributions described in Fig.~\ref{G0_250_var_NATOMS} and Fig.~\ref{G0_250_DDI_0.0_var_NATOMS}. 

\begin{figure}[h]
	\centering
	\includegraphics[width = 0.95\linewidth]{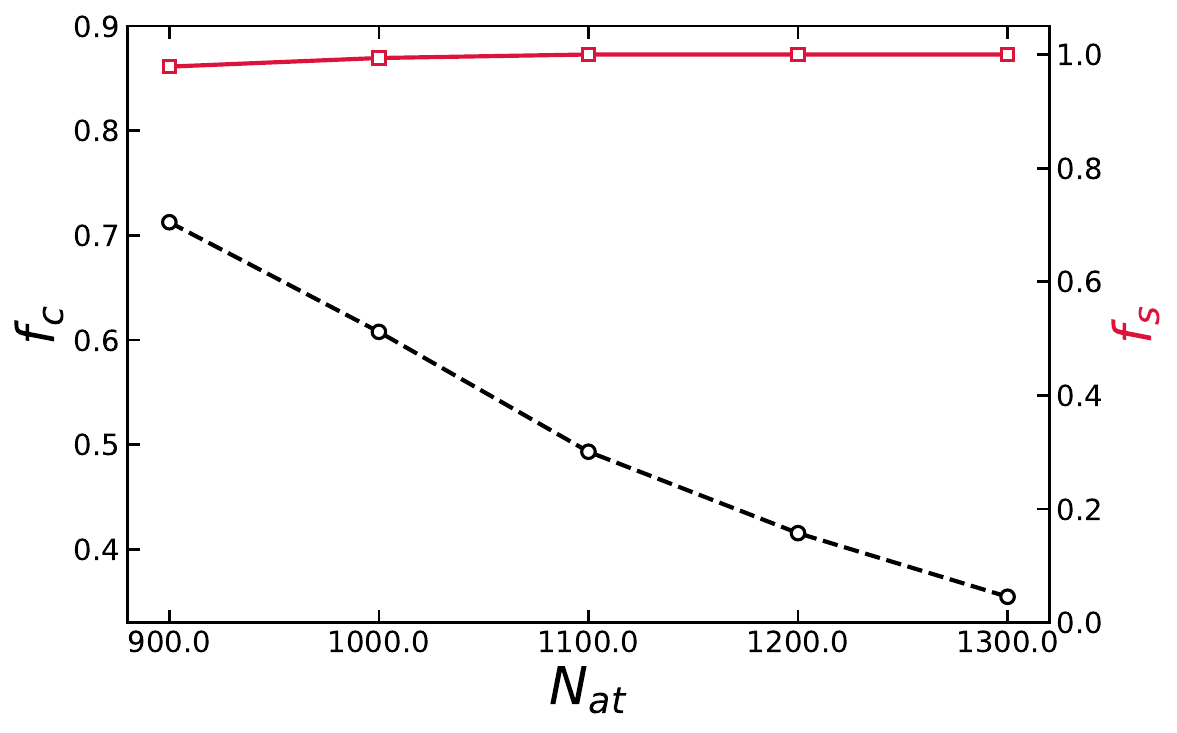}
	\caption{(Color online) Condensate fraction $f_c$ (dashed black line) and superfluid fraction $f_s$ (solid red line) versus particle number $N_{at}$ for the parameters corresponding to Fig.~\ref{G0_250_var_NATOMS} are described here. The $y$-axis describes condensate fraction while the right $y$-axis denotes the superfluid fraction.} 
	\label{fig:G0_250}
\end{figure}

\begin{figure}[h]
	\centering
	\includegraphics[width = 0.95\linewidth]{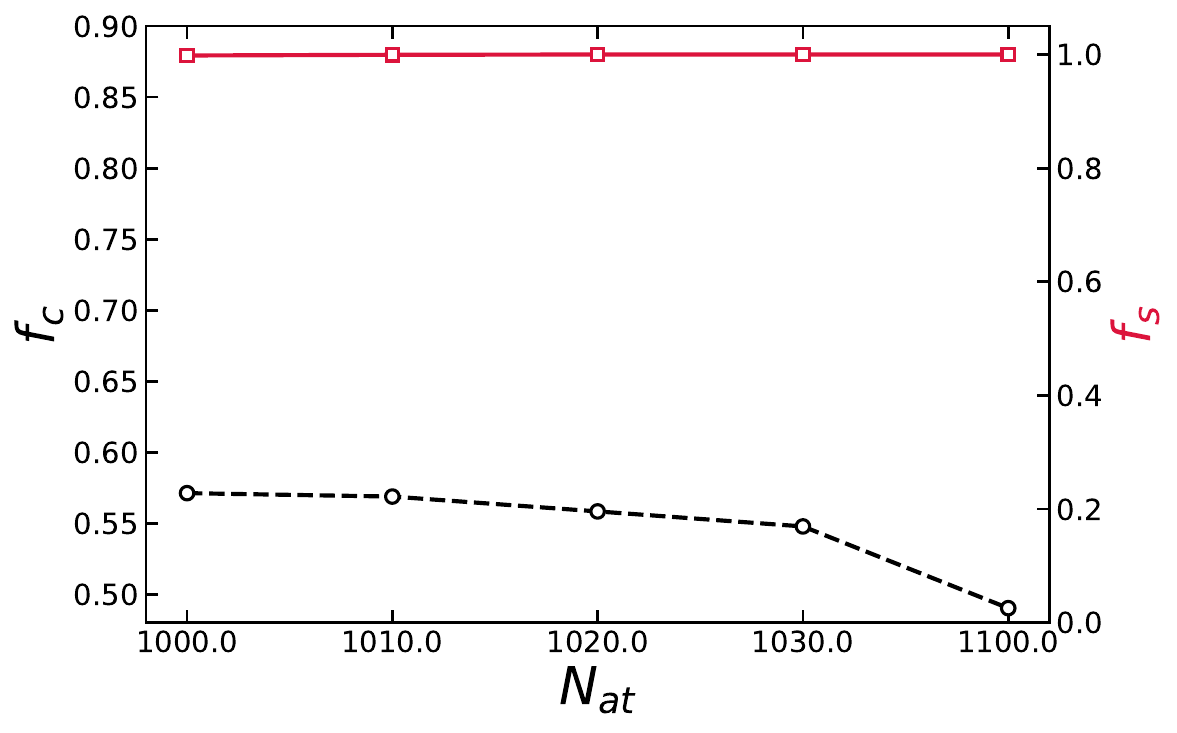}
	\caption{(Color online) Particle-number dependence of the condensate fraction $f_c$ (dashed black line) and the superfluid fraction $f_s$ (solid red line) for the parameters corresponding to Fig.~\ref{G0_250_DDI_0.0_var_NATOMS}. The results demonstrate the evolution of condensate and superfluid properties with increasing particle number in the absence of the dipole-dipole interaction.
}
	\label{fig:G0_250_DDI0}
\end{figure}

However, a closer inspection of the modulational instability, depicted in Fig.~\ref{G0_250_var_NATOMS}($b$) and Fig~\ref{G0_250_DDI_0.0_var_NATOMS}($b$) reveals that the area of the side lobes varies differently in presence and absence of dipolar interaction. The area of the side lobes decreases with increasing particle number (when DDI is present), while in the absence of dipolar potential, the area increases. The latter is the more conventional situation as the MI gain factor ($\Gamma$) is proportional to the square root of the effective interaction strength ($g_{eff}$, which takes into account all the interactions applied on the system) and the particle number ($\Gamma\propto\sqrt{g_{eff}N_{at}}$). However, the counter-intuitive result in Fig.~\ref{G0_250_var_NATOMS}($b$) where the side lobe area is decreasing with increasing particle number is a result of the competition between the attractive MF, BMF and 3B interaction and the repulsive DDI. The repulsive DDI is opposing the growth of the modulational instability as the effective interaction is weaker (due to competition between MF, BMF, 3B and DDI) even though the particle number has increased. Nevertheless, this competion also results in considerable depletion of the condensate fraction ($\sim42\%$) as noted in Fig.~\ref{fig:G0_250} while the superfluid fraction remains pinned at around $1$. 
In absence of DDI, we also observe depletion of $f_c$, however it is not as pronounced as the previous which is about $14\%$ (see Fig.~\ref{fig:G0_250_DDI0}).

%Another case that we have taken DDI is zero and 2B interaction strength is $250g_{0}$, 3B and BMF strength attractive. So by taking dipole dipole interaction zero we have observed the gray droplet solution for the number of atoms 1000. And, further increasing the $N_{at}$ it nature of density solution converts into flat-top with lower density value (Fig.\ref{G0_250_DDI_0.0_var_NATOMS}(a)) which is also same till $N_{at}$ is 1500, see Fig.\ref{G0_250_DDI_0.0_var_NATOMS}(b). 

 \begin{comment}
 \begin{figure}[h]
	\centering
	\includegraphics[width = 0.95\linewidth]{energy_rms_900-1300.pdf}
	\caption{Here value of 2B strength considered as $250 g_{0}$. Therefore, right figures describe the behavior of energy per particle while left is the explanation of droplet size by varying particle number from $900$ to $1300$. Accounting strength values: $g_{0} = 75.1431 $, $g' = 161.7409$, $g_{3B} = -1.57577 \times 10^{-5}$ and $g_{dd} = 16.82783$}
	\label{fig:energy_rms_900-1300_Energy_RMS}
\end{figure}
\begin{figure}[h]
	\centering
	\includegraphics[width = 0.95\linewidth]{energy_rms-1000-1100_DDI0.pdf}
	\caption{Value of 2B strength considered as $250 g_{0}$ while $g_{dd} = 0.0$. Therefore, right figures describe the behavior of energy per particle while left is the explanation of droplet size with varying particle number from $900$ to $1300$. Accounting strength values: $g_{0} = 75.1431 $, $g' = 161.7409$, $g_{3B} = -1.57577 \times 10^{-5}$.}
	\label{fig:energy_rms_1000-1100_Energy_RMS}
\end{figure}
\end{comment}

\section{Summary and Conclusion}\label{conclusion}
In this article, we have investigated the fragmentation of the quantum fluid due to the competition between two body mean-field (MF), three body (3B), Beyond mean field (BMF) and dipole-dipole interactions (DDI). Specifically, we concentrated on the MF and DDI interaction with non zero BMF interaction. We have observed that the presence of BMF interaction is essential to observe the fragmentation of the fluid. The physical system under consideration is a quasi one dimensional homogeneous condensate. The fragmentation is analyzed in the light of dispersion mechanism. Additionally, we calculate the condensate fraction and superfluid fraction with changing Hamiltonian parameters. 

First we report the fragmentation of localized condensate where its dispersion curve clearly demonstrates roton mode, suggesting crystallization of the condensate resulting in depletion of the condensate fraction and strengthening the superfluid fraction leading to the possible onset of supersolidity.  

Next we investigate the quantum droplets with flat-top density distribution. The calculation is carried out for fixed DDI and varying the MF interaction. The fragmentation of the droplets is observed for stronger MF interaction. The dispersion mechanism shows the emergence of modulational instability while the side lobe area increases with stronger interaction resulting in the depletion of the condensate. The superfluid fraction remains relatively fixed. 

When we study the fragmentation of the droplets as a function of particle number, again we observe modulational instability. In the presence of attractive MF, BMF and 3B interaction and repulsive DDI, a sharp drop in condensate fraction is observed as a function of particle number while switching off the DDI results a smother fall of $f_c$. In both cases the superfluid fraction does not change significantly. This is due to the fact that the quantum fluctuations deplete the zero-momentum state, while superfluidity is primarily dictated by the system's overall phase coherence and density correlations. In the overall calculation we are unable to see any significant contribution of the three-body interaction this might be due to its weak presence in a more dominating MF and DDI interaction landscape.

%sIn all the observations the 3B interaction played a more inert role however, we have carried out our calculation with 3B interaction for the sake of completeness. 

We hope our current analysis will play an important role in understanding the supersolid phase and quantum droplets and their inherent physical nature and will lead to experimental verification.
\section*{Acknowledgement}
AK thanks the Council of Scientific and
Industrial Research (CSIR) Human Resource
Development Group (HRDG) Extramural Research Division (EMR-II), India for the
support provided through project number
03/1500/23/EMR-II.

\newpage
\bibliographystyle{apsrev4-1}
\bibliography{ms_v1}

\end{document}